\documentclass[12pt]{article}
\usepackage{axodraw,bbold}

\catcode`@=12
\begin{document}
\thispagestyle{empty}
\begin{flushright} 
UCRHEP-T431\\ 
June 2007\
\end{flushright}
\vspace{0.5in}
\begin{center}
{\Large	\bf Non-Abelian Discrete Flavor Symmetries\\}
\vspace{1.5in}
{\bf Ernest Ma\\}
\vspace{0.2in}
{\sl Department of Physics and Astronomy, University of 
California,\\ Riverside, California 92521, USA\\}
\vspace{1.0in}
\end{center}

\begin{abstract}\
This is an incomplete survey of some non-Abelian discrete symmetries 
which have been used recently in attempts to understand the flavor 
structure of leptons and quarks.  To support such symmetries, new scalar 
particles are required.  In some models, they are very massive, in which 
case there may not be much of a trace of their existence at the TeV scale.   
In other models, they are themselves at the TeV scale, in which case there 
is a reasonable chance for them to be revealed at the LHC (Large Hadron 
Collider) at CERN. 
\end{abstract}

\newpage
\baselineskip 24pt

\section*{Introduction}

Leptons and quarks come in three families.  Their flavor structure, i.e. the 
specific values of their mass and mixing matrices, has long been a puzzle 
and a subject of study.  In recent years, with the steady accumulation of 
data on solar and atmospheric neutrino oscillations, the lepton mixing 
matrix has been determined to a large extent and it came as a surprise 
to many that it does not resemble at all the known quark mixing matrix. 
Is there a way to understand this?  One possible approach is the use 
of non-Abelian discrete symmetries, such as $S_3$ and $A_4$ among others. 
In this report, I will survey this topic, offering a basic recipe for 
constructing models, with an extensive (but nevertheless incomplete) 
list of references.

\section*{Finite Groups}

To obtain a non-Abelian discrete symmetry, a simple heuristic way is to 
choose two specific noncommuting matrices and form all possible products. 
As a first example, consider the two $2 \times 2$ matrices:
\begin{equation}
A = \pmatrix{0 & 1 \cr 1 & 0}, ~~~ B = \pmatrix{\omega & 0 \cr 0 & 
\omega^{-1}},
\end{equation}
where $\omega^n = 1$, i.e. $\omega = \exp(2\pi i/n)$.  Since $A^2=1$ and 
$B^n=1$, this group contains $Z_2$ and $Z_n$.  For $n=1,2$, we obtain 
$Z_2$ and $Z_2 \times Z_2$ respectively, which are Abelian.  For $n=3$, 
the group generated has 6 elements and is in fact the smallest non-Abelian 
finite group $S_3$, the permutation group of 3 objects.  This particular 
representation is not the one found in text books, but is related to it 
by a unitary transformation \cite{Ma:2004pt}, and was first used in 1990 for 
a model of quark mass matrices \cite{Ma:1990qh,Deshpande:1991zh}. For $n=4$, 
the group generated has 8 elements which are in fact $\pm 1$, 
$\pm \sigma_1$, $\pm \sigma_2$, $\pm i \sigma_3$, 
where $\sigma_{1,2,3}$ are the usual Pauli spin matrices.  This forms the 
group $D_4$, i.e. the symmetry group of the square, which was 
first used in 2003 \cite{Grimus:2003kq,Grimus:2004rj}.  If the 8 elements 
are $\pm 1$, $\pm i\sigma_{1,2,3}$ instead, then they form the group 
of quaternions $Q$, which has also been used \cite{Frigerio:2004jg} 
for quark and lepton mass matrices.  In general, the groups generated by 
Eq.~(1) have $2n$ elements and may be denoted as $\Delta(2n)$.

Consider next the two $3 \times 3$ matrices:
\begin{equation}
A = \pmatrix{0 & 1 & 0 \cr 0 & 0 & 1 \cr 1 & 0 & 0}, ~~~ 
B = \pmatrix{\omega & 0 & 0 \cr 0 & \omega^2 & 0 \cr 0 & 0 & \omega^{-3}}. 
\end{equation}
Since $A^3=1$ and $B^n=1$, this group contains $Z_3$ and $Z_n$.  For $n=1$, 
we obtain $Z_3$.  For $n=2$, the group generated has 12 elements and is 
$A_4$, the even permutation group of 4 objects, which was first used in 
2001 in a model of lepton mass matrices \cite{Ma:2001dn,Babu:2002dz}.  It 
is also 
the symmetry group of the tetrahedron, one of five perfect geometric 
solids, identified by Plato with the element ``fire'' \cite{Ma:2002ge}.  In 
general, the groups generated by Eq.~(2) have $3n^2$ elements and may 
be denoted as $\Delta(3n^2)$ \cite{Luhn:2007uq}.  They are in fact subgroups 
of 
$SU(3)$. In particular, $\Delta(27)$ has also been used \cite{Ma:2006ip,
deMedeirosVarzielas:2006fc}. 
Generalizing to $k \times k$ matrices, we then have the series 
$\Delta(kn^{k-1})$. However, since there are presumably only 3 families, 
$k>3$ is probably not of much interest.

Going back to $k=2$, but using instead the following two matrices:
\begin{equation}
A = \pmatrix{0 & 1 \cr 1 & 0}, ~~~ B = \pmatrix{\omega & 0 \cr 0 & 1}.
\end{equation}
Now again $A^2=1$ and $B^n=1$, but the group generated will have $2n^2$ 
elements.  Call it $\Sigma(2n^2)$.  For $n=1$, it is just $Z_2$.  For 
$n=2$, it is $D_4$ again.   For $k=3$, consider
\begin{equation}
A = \pmatrix{0 & 1 & 0 \cr 0 & 0 & 1 \cr 1 & 0 & 0}, ~~~ 
B = \pmatrix{\omega & 0 & 0 \cr 0 & 1 & 0 \cr 0 & 0 & 1}, 
\end{equation}
then the groups generated have $3n^3$ elements and may be denoted as 
$\Sigma(3n^3)$. They are in fact subgroups of $U(3)$.  For $n=1$, it is 
just $Z_3$.  For $n=2$, it is $A_4 \times Z_2$.  For $n=3$, the group 
$\Sigma(81)$ has been used \cite{Ma:2006ht} to understand the Koide formula 
\cite{Koide:1982si} as well as lepton mass matrices \cite{Ma:2007ku}.  
In general, we have the series $\Sigma(kn^k)$.

\section*{Model Recipe} 

\noindent (I) Choose a group, e.g. $S_3$ or $A_4$, and write down its 
possible representations.  For example $S_3$ has \underline{1}, 
\underline{1}$'$, \underline{2}; $A_4$ has \underline{1}, \underline{1}$'$, 
\underline{1}$''$, \underline{3}.  Work out all product decompositions. 
For example $\underline{2} \times \underline{2} = \underline{1} + 
\underline{1}' + \underline{2}$ in $S_3$, and $\underline{3} \times 
\underline{3} = \underline{1} + \underline{1}' + \underline{1}'' + 
\underline{3} + \underline{3}$ in $A_4$.

\noindent (II) Assign $(\nu,l)_{1,2,3}$ and $l^c_{1,2,3}$ to the 
representations of your choice.  If you want to consider only 
renormalizable interactions, you will need to add Higgs doublets 
(and perhaps also triplets and singlets) and assign them.  You may 
also want to consider adding neutrino singlets.

\noindent (III) Because of your choice of particle content and their 
representations, the Yukawa structure of your model is restricted. 
As the Higgs bosons acquire vacuum expectation values (which may be related 
because of some extra or residual symmetry), your lepton mass matrices 
will have certain particular forms.  If the number of parameters 
involved are less than the number of observables, you then have one 
or more predictions.  Of course, the forms themselves have to be 
consistent with the known values of $m_e$, $m_\mu$, $m_\tau$, etc.

\noindent (IV) Because there will be more than one Higgs doublet in 
such models, flavor nonconservation will appear at some level.  You will 
need to work out its phenomenological consequences, making sure that 
your model is consistent with present experimental constraints.  You 
can then proceed to explore its observability at the TeV scale.

\noindent (V) If you insist on using only the one Higgs doublet of the 
Standard Model, then you must consider effective nonrenormalizable 
interactions to support the discrete flavor symmetry.  In such models, 
there are no predictions beyond the forms of the mass matrices themselves.

\noindent (VI) Quarks can be considered in the same way.  The two quark 
mass matrices ${\cal M}_u$ and ${\cal M}_d$ must be nearly aligned so that 
their mixing matrix involves only small angles.  In contrast, ${\cal M}_\nu$ 
and ${\cal M}_l$ should have different structures so that large angles 
can be obtained.

\section*{$S_3$}

Being the simplest, the non-Abelian discrete symmetry $S_3$ was used 
already \cite{Yamaguchi:1964} in the early days of strong interactions.  There 
are many recent applications \cite{Kubo:2003iw,Kubo:2004ps,Chen:2004rr,
Grimus:2005mu,Lavoura:2005kx,Teshima:2005bk,
Koide:2005ep,Mohapatra:2006pu,Morisi:2005fy,Kaneko:2006wi}, some of which are 
discussed in my talk 
at VI-Silafae \cite{Ma:2006ay}.  Typically, such models often require 
extra symmetries beyond $S_3$ to reduce the number of parameters, 
or assumptions of how $S_3$ is spontaneously and softly broken. 
For illustration, consider the model of Kubo et al.~\cite{Kubo:2003iw} 
which has recently been updated by Felix et al.~\cite{Felix:2006pn}. 
The symmetry used is actually $S_3 \times Z_2$, with the assignments
\begin{equation}
(\nu,l), ~l^c, ~N, ~(\phi^+,\phi^0) \sim \underline{1} + \underline{2},
\end{equation}
and equal vacuum expectation values for the two Higgs doublets transforming 
as \underline{2} under $S_3$.  The $Z_2$ symmetry serves to eliminate 4 
Yukawa couplings otherwise allowed by $S_3$, resulting in an inverted 
ordering of neutrino masses with
\begin{equation}
\theta_{23} \simeq \pi/4, ~~ \theta_{13} \simeq 0.0034, ~~ 
m_{ee} \simeq 0.05~{\rm eV},
\end{equation}
where $m_{ee}$ is the effective Majorana neutrino mass measured in 
neutrinoless double beta decay.  This model relates $\theta_{13}$ to the 
ratio $m_e/m_\mu$.

\section*{$A_4$}

To understand why quarks and leptons have very different mixing matrices, 
$A_4$ turns out to be very useful.  It allows the two different quark mass 
matrices to be diagonalized by the same unitary tranformations, implying 
thus no mixing as a first approximation, but because of the assumed 
Majorana nature of the neutrinos, a large mismatch may occur in the lepton 
sector, thus offering the possibility of obtaining the so-called 
tribimaximal mixing matrix \cite{Harrison:2002er,He:2003rm}, which is a good 
approximation to present data. One way of doing this is to consider 
the decomposition
\begin{equation}
U_{l\nu} = \pmatrix{\sqrt{2/3} & 1/\sqrt{3} & 0 \cr -1/\sqrt{6} & 1/\sqrt{3} 
& -1/\sqrt{2} \cr -1/\sqrt{6} & 1/\sqrt{3} & 1/\sqrt{2}} = 
{1 \over \sqrt{3}} \pmatrix{1 & 1 & 1 \cr 1 & \omega & \omega^2 \cr 1 & 
\omega^2 & \omega} \pmatrix{0 & 1 & 0 \cr 1/\sqrt{2} & 0 & -i/\sqrt{2} \cr 
1/\sqrt{2} & 0 & i/\sqrt{2}},
\end{equation}
where $\omega = \exp(2 \pi i/3) = -1/2 + i\sqrt{3}/2$.  The matrix involving 
$\omega$ has equal moduli for all its entries and was conjectured already 
in 1978 \cite{Cabibbo:1977nk,Wolfenstein:1978uw} to be a possible candidate 
for the $3 \times 3$ 
neutrino mixing matrix.

Since $U_{l\nu} = U_l^\dagger U_\nu$, where $U_l$, $U_\nu$ diagonalize 
${\cal M}_l {\cal M}_l^\dagger$, ${\cal M}_\nu {\cal M}_\nu^\dagger$ 
respectively, Eq.~(7) may be obtained if we have
\begin{equation}
U_l^\dagger =  {1 \over \sqrt{3}} \pmatrix{1 & 1 & 1 \cr 1 & \omega & 
\omega^2 \cr 1 & \omega^2 & \omega}
\end{equation}
and
\begin{equation}
{\cal M}_\nu = \pmatrix{a+2b & 0 & 0 
\cr 0 & a-b & d \cr 0 & d & a-b}
\end{equation}
\begin{equation} 
= \pmatrix{0 & 1 & 0 \cr 1/\sqrt{2} & 0 & -i/\sqrt{2} \cr 1/\sqrt{2} 
& 0 & i/\sqrt{2}} \pmatrix{a-b+d & 0 & 0 \cr 0 & a+2b & 0 \cr 0 & 0 & 
-a+b+d} \pmatrix{0 & 1/\sqrt{2} & 1/\sqrt{2} \cr 1 & 0 & 0 \cr 
0 & -i/\sqrt{2} & i/\sqrt{2}}.
\end{equation}

It was discovered in Ref.~\cite{Ma:2001dn} that Eq.~(8) is naturally obtained 
with $A_4$ if
\begin{equation}
(\nu,l)_{1,2,3} \sim \underline{3}, ~~ l^c_{1,2,3} \sim \underline{1} + 
\underline{1}' + \underline{1}'', ~~  (\phi^+,\phi^0)_{1,2,3} \sim 
\underline{3}
\end{equation}
for $\langle \phi^0_1 \rangle = \langle \phi^0_2 \rangle = \langle \phi^0_3 
\rangle$.  This assignment also allows $m_e$, $m_\mu$, $m_\tau$ to take 
on arbitrary values because there are here exactly three independent 
Yukawa couplings invariant under $A_4$.  If we use this also for quarks 
\cite{Babu:2002dz}, then $U_u^\dagger$ and $U_d^\dagger$ are also given by 
Eq.~(8), 
resulting in $V_{CKM}=1$, i.e. no mixing.  This should be considered as a 
good first approximation because the observed mixing angles are all small. 
In the general case without any symmetry, we would have expected $U_u$ and 
$U_d$ to be very different.

It was later discovered in Ref.~\cite{Ma:2004zv} that Eq.~(9) may also be 
obtained with $A_4$, using two further assumptions.  Consider the 
most general $3 \times 3$ Majorana mass matrix in the form
\begin{equation}
{\cal M}_\nu = \pmatrix{a+b+c & f & e \cr f & a+b\omega +c\omega^2 & d \cr 
e & d & a+b\omega^2 +c\omega},
\end{equation}
where $a$ comes from \underline{1}, $b$ from \underline{1}$'$, $c$ from 
\underline{1}$''$, and $(d,e,f)$ from \underline{3} of $A_4$.  To get 
Eq.~(9), we need $e=f=0$, i.e. the effective scalar $A_4$ triplet 
responsible for neutrino masses should have its vacuum expectation value 
along the (1,0,0) direction, whereas that responsible for charged-lepton 
masses should be (1,1,1) as I remarked earlier.  This misalignment is a 
technical challenge to all such models \cite{Altarelli:2005yp,Babu:2005se,
Ma:2005sh,Zee:2005ut,Ma:2005qf,Altarelli:2005yx,
He:2006dk,Adhikary:2006wi,Adhikary:2006jx,Yin:2007rv,Altarelli:2006kg,
He:2006et}.  The other requirement is that 
$b=c$. Since they come from different representations of $A_4$, this is rather 
{\it ad hoc}.  A very clever solution \cite{Altarelli:2005yp,Altarelli:2005yx} 
is to eliminate both, 
i.e. $b=c=0$.  This results in a normal ordering of neutrino masses with 
the prediction 
\cite{Ma:2005sh}
\begin{equation}
|m_{\nu_e}|^2 \simeq |m_{ee}|^2 + \Delta m^2_{atm}/9.
\end{equation}
Other applications \cite{Ma:2002iq,Ma:2002yp,Babu:2002ki,Chen:2005jm,
Hirsch:2005mc,
Ma:2005qy,Ma:2005mw,Ma:2005tr,
Ma:2006sk,Ma:2006wm,Ma:2006vq,Lavoura:2006hb,deMedeirosVarzielas:2005qg,
King:2006np,Morisi:2007ft,Koide:2007kw,Hirsch:2007kh} of $A_4$ have also been 
considered.  A natural (spinorial) extension of $A_4$ is the binary 
tetrahedral group \cite{Aranda:1999kc,Aranda:2000tm} which is under active 
current discussion \cite{Carr:2007qw,Feruglio:2007uu,Chen:2007af,
Frampton:2007et}.

\section*{Others}

Other recent applications of non-Abelian discrete flavor symmetries include 
those of $D_4$ \cite{Grimus:2003kq,Grimus:2004rj,Kobayashi:2006wq}, 
$Q_4$ \cite{Frigerio:2004jg}, $D_5$ 
\cite{Ma:2004br,Hagedorn:2006ir}, $D_6$ \cite{Kajiyama:2006ww}, $Q_6$ 
\cite{Frampton:1994xm,Babu:2004tn,Kubo:2005ty}, $D_7$ 
\cite{Chen:2005jt}, $S_4$ \cite{Ma:2005pd,Hagedorn:2006ug,Cai:2006mf,
Zhang:2006fv,Koide:2007sr},  $\Delta(27)$ \cite{Ma:2006ip,
deMedeirosVarzielas:2006fc}, $\Delta(75)$ \cite{Kaplan:1993ej,
Schmaltz:1994ws}, $\Sigma(81)$ \cite{Ma:2006ht,
Ma:2007ku}, 
and $B_3 \times Z_2^3$ \cite{Grimus:2005rf,Grimus:2006wy} which has 384 
elements.

This work was supported in part by the U.~S.~Department of Energy under Grant 
No.~DE-FG03-94ER40837.

\newpage

\baselineskip 18pt

\bibliographystyle{unsrt}

\end{document}